\documentclass[conference]{IEEEtran}

\usepackage{graphics,graphicx,epsfig,color,ulem}
\usepackage{subfigure}  

\usepackage{amsfonts,amsmath,amssymb}   

\usepackage{flushend}

\begin{document}

\title{Testing the Gravitational Redshift \\with Atomic 
Gravimeters?}

\author{\authorblockN{Peter Wolf$^1$, Luc Blanchet$^2$, Christian
    J. Bord\'{e}$^{1,3}$, \\Serge Reynaud$^4$, Christophe Salomon$^5$,
    Claude Cohen-Tannoudji$^5$} \authorblockA{\\ $^1$LNE-SYRTE,
    Observatoire de Paris, CNRS, UPMC, France \\ $^2$GRECO, Institut
    d'Astrophysique de Paris, CNRS, UPMC, France \\ $^3$Laboratoire de
    Physique des Lasers, Universit\'{e} Paris 13, CNRS, France \\
    $^4$Laboratoire Kastler Brossel, CNRS, ENS, UPMC, France \\
    $^5$Laboratoire Kastler Brossel et Coll\`ege de France, CNRS, ENS,
    UPMC, France}}


%


\maketitle

\begin{abstract}
  Atom interferometers allow the measurement of the acceleration of
  freely falling atoms with respect to an experimental platform at
  rest on Earth's surface. Such experiments have been used to test the
  universality of free fall by comparing the acceleration of the atoms
  to that of a classical freely falling object. In a recent paper,
  M\"uller, Peters and Chu [Nature {\bf 463}, 926-929 (2010)] argued
  that atom interferometers also provide a very accurate test of the
  gravitational redshift (or universality of clock rates). Considering
  the atom as a clock operating at the Compton frequency associated
  with the rest mass, they claimed that the interferometer measures
  the gravitational redshift between the atom-clocks in the two paths
  of the interferometer at different values of gravitational
  potentials. In the present paper we analyze this claim in the frame
  of general relativity and of different alternative theories, and
  conclude that the interpretation of atom interferometers as testing
  the gravitational redshift at the Compton frequency is unsound.  The
  present work is a summary of our extensive paper [Wolf \textit{et
    al}., arXiv:1012.1194, Class. Quant. Grav. in press], to which the
  reader is referred for more details.
\end{abstract}


%
\IEEEpeerreviewmaketitle

\section{Introduction}\label{I}

The gravitational redshift, considered as one of the ``classical
tests'' of general relativity (GR), is actually a test of one facet of
the Einstein equivalence principle (EEP). The weak version of the
equivalence principle has been verified with high precision using
torsion balances \cite{Schlamminger} and Lunar laser ranging
\cite{Williams}. The gravitational redshift was predicted by Einstein
but not observed for a long time. It became observable with the advent
of high precision quantum spectroscopy and was first measured in 1960
by Pound and Rebka \cite{Pound}, who used gamma ray spectroscopy of
the radiation emitted and absorbed by ${}^{57}$Fe nuclei. The emitter
and absorber were placed at the top and bottom of a $22.5\,\mathrm{m}$
high tower at Harvard and the frequency difference predicted by GR was
measured with about $1\%$ uncertainty.

In a test of the gravitational redshift using clocks, one checks
that the clock rates are universal --- i.e. the relative rates
depend only on the difference of gravitational potentials (as
determined by the trajectories of massive test bodies) but not on
the nature and internal structure of the clocks. In 1976 a
hydrogen-maser clock was launched on a rocket to an altitude of
$10,000\,\mathrm{km}$ and its frequency compared to a similar
clock on ground. This yielded a test of the gravitational redshift
with about $10^{-4}$ accuracy \cite{Vessot}. The European Space
Agency will fly in 2013 the Atomic Clock Ensemble in Space (ACES)
\cite{Cacciapuoti}, including a highly stable laser-cooled atomic
clock, which will test (in addition to many other applications in
fundamental physics and metrology) the gravitational redshift to a
precision of about $10^{-6}$. The gravitational redshift is also
tested in null redshift experiments in which the rates of
different clocks (based on different physical processes or
different atoms) are compared to each other \cite{Blatt}.

Quite generally in a modern context, tests of GR measure the
difference between the predictions of GR and of some generalized
alternative theory or theoretical framework (see \cite{Will} for a
review). It has to be kept in mind that the classification and
inter-comparison of different tests have then to be defined with
respect to the framework used.

Atom interferometers have reached high sensitivities in the
measurement of the gravitational acceleration
\cite{Peters,refsyrte}. This yields very important tests of the weak
equivalence principle (WEP) or universality of free fall (UFF) when
comparing the free fall of atoms with that of classical macroscopic
matter (in practice a nearby freely falling corner cube whose
trajectory is monitored by lasers). The relative precision of such
tests of the UFF is currently $7\times10^{-9}$, using Cs
\cite{Peters,Mueller} or Rb \cite{refsyrte} atoms.  Although it
remains less sensitive than tests using macroscopic bodies of
different composition \cite{Williams,Schlamminger} which have reached
a precision of $2\times10^{-13}$, this UFF test is interesting as it
is the most sensitive one comparing the free fall of quantum objects
(namely C{\ae}sium atoms) with that of a classical test mass (the
corner cube). In this contribution, which summarizes the extensive
paper \cite{WolfCQG}, we investigate whether such experiments can also
be interpreted as tests of the gravitational redshift.

\section{Overview of our extensive paper \cite{WolfCQG}
}\label{IB}

In a recent paper, M\"uller, Peters and Chu \cite{Mueller}
(hereafter abbreviated as MPC) proposed a new interpretation of
atom interferometry experiments as testing the gravitational
redshift, that is also the universality of clock rates (UCR), with
a precision $7\times10^{-9}$, which is several orders of magnitude
better than the best present \cite{Vessot} and near future
\cite{Cacciapuoti} clock tests.

\subsection{Analogy with clock experiments}

The main argument of MPC (see also the more detailed papers
\cite{Hohensee,Hohensee2}) is based on an analogy between atom
interferometry experiments and classical clock experiments. The
idea of clock experiments is to synchronize a pair of clocks when
they are located closely to one another, and move them to
different elevations in a gravitational field. The gravitational
redshift will decrease the oscillation frequency of the lower
clock relative to the higher one, yielding a measurable phase
shift between them. There are two methods for measuring the
effect. Either we bring the clocks back together and compare the
number of elapsed oscillations, or we measure the redshift by
means of continuous exchanges of electromagnetic signals between
the two clocks. In both methods one has to monitor carefully the
trajectories of the two clocks. For example, in the second method
one has to remove the Doppler shifts necessarily appearing in the
exchanges of electromagnetic signals.

In the first method, the phase difference between the two clocks
when they are recombined together, can be written as a difference
of integrals over proper time,
\begin{equation}\label{phase_clock}
  \Delta \varphi_{\textrm{clock}} = \omega \left[\int_{\textrm{I}} \textrm{d}\tau -
  \int_{\textrm{II}} \textrm{d}\tau \right] \equiv \omega \oint \textrm{d}\tau\,.
\end{equation}
The two clocks have identical proper frequency $\omega$. We denote
by I and II the two paths (with say I being at a higher altitude,
i.e. a lower gravitational potential) and use the notation
$\oint\textrm{d}\tau$ to mean the difference of proper times
between the two paths, assumed to form a close contour. The
integrals in \eqref{phase_clock} are evaluated along the paths of
the clocks, and we may use the Schwarzschild metric to obtain an
explicit expression of the measured phase shift in the
gravitational field of the Earth.

\begin{figure}
\begin{center}
\includegraphics[width=9cm,angle=0]{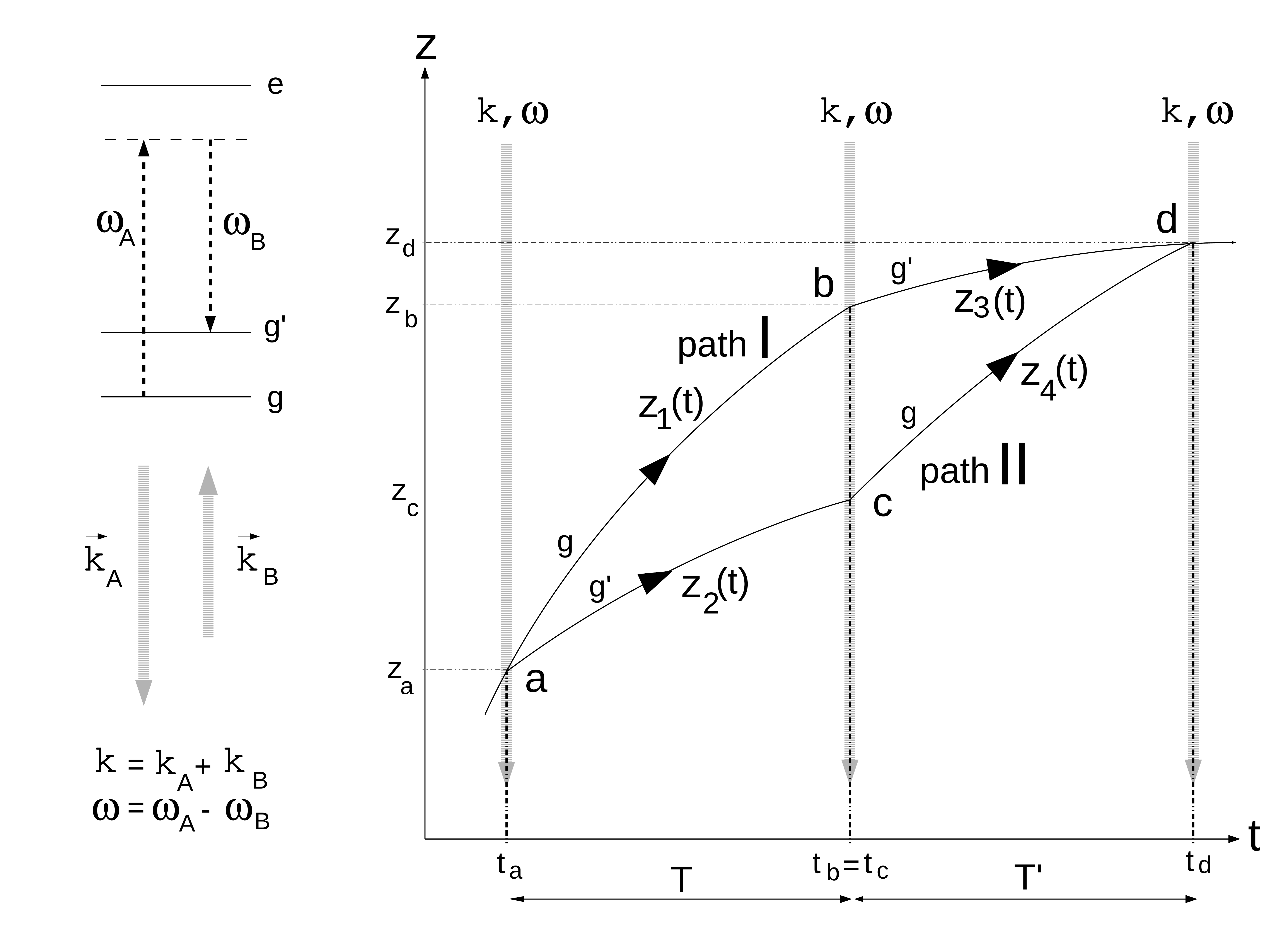}
\caption{Space-time trajectories followed by the atoms in the
  interferometer. Laser pulses occur at times $t_a$, $t_b=t_c$ and
    $t_d$, separated by time intervals $T$ and $T'$ (we have $T'=T$ when gravity gradients are neglected). The two-photon
    Raman transitions are between two hyperfine levels $g$ and $g'$ of
    the ground state of an alkali atom.}
\label{fig1}
\end{center}
\end{figure}

The phase shift measured by an atom interferometer contains a
contribution which is similar to the clock phase shift
\eqref{phase_clock}, so it is tempting to draw an analogy with clock
experiments. In this analogy, the role of the clock's proper frequency
is played by the atom's (de Broglie-)Compton frequency
$\omega_\textrm{C} = {m c^2/\hbar}$, where $m$ denotes the rest mass
of the atom. However the phase shift includes also another
contribution $\Delta \varphi_\ell$ coming from the interaction of the
laser light used in the beam-splitting process with the atoms. Thus,
\begin{equation}\label{phase_interf}
\Delta \varphi = \omega_\textrm{C} \oint \textrm{d}\tau + \Delta
\varphi_\ell\,.
\end{equation}
Here we are assuming that the two paths close up at the entry and
exit of the interferometer; otherwise, additional terms have to be
added to \eqref{phase_interf}. The schematic view of the atom
interferometer showing the two interferometer paths I and II is given
by Fig.~\ref{fig1}. The first term in
\eqref{phase_interf} is proportional to the atom's mass through
the Compton frequency and represents the difference of Compton
phases along the two classical paths. In contrast, the second term
$\Delta \varphi_\ell$ does not depend on the mass of the atoms.

At first sight, the first term in \eqref{phase_interf} could be
used for a test of the gravitational redshift, in analogy with
classical clock experiments, and the
precision of the test could be very good, because the Compton
frequency of the C{\ae}sium atom is very high, $\omega_\textrm{C}
\approx 2 \pi \times 3.0 \times 10^{25}\,\textrm{Hz}$. However, as
shown in a previous brief comment \cite{WolfNat} and the detailed
paper \cite{WolfCQG}, this re-interpretion of the atom interferometer
as testing the UCR is fundamentally incorrect.\footnote{See also the
  reply of MPC \cite{MuellerNat} to our brief comment. After
  completion of our work \cite{WolfNat,WolfCQG}, several independent
  analysis have appeared \cite{Samuel,Giulini} supporting our views
  and consistent with our conclusions.}

At this point, we want to mention other crucial differences between
atom interferometry and clock experiments, which reinforce our
conclusions. In clock experiments the trajectories of the clocks are
continuously controlled for instance by continuous exchange of
electromagnetic signals. In atom interferometry in contrast, the
trajectories of the atoms are not measured independently but
theoretically derived from the Lagrangian and initial
conditions. Furthermore, we expect that it is impossible to determine
independently the trajectories of the wave packets without destroying
the interference pattern at the exit of the interferometer.

Another important difference lies in the very notion of a clock.
Atomic clocks use the extremely stable energy difference between two
internal states. By varying the frequency of an interrogation signal
(e.g. microwave or optical), one obtains a resonant signal when the
frequency is tuned to the frequency of the atomic transition. In their
re-interpretation, MPC view the entire atom as a clock ticking at the
Compton frequency associated with its rest mass. But the
``atom-clock'' is not a real clock in the previous sense, since it
does not deliver a physical signal at Compton frequency, as also
recently emphasized in the same context in \cite{Samuel}.

Note also that the phase shift
\eqref{phase_clock} for clocks is valid in any gravitational field,
with any gravity gradients, since it is simply the proper time
elapsed along the trajectories of the clocks in a gravitational
field. By contrast, the phase shift \eqref{phase_interf} is known
only for quadratic Lagrangians \cite{Storey} and cannot be
applied in a gravitational field with large gravity gradients, or
more generally with any Lagrangian that is of higher order.

\subsection{Analysis in general relativity}

The clear-cut argument showing that the atom interferometer does not
measure the redshift is that the ``atom-clock'' contribution, i.e.
the first term in \eqref{phase_interf}, is in fact \textit{zero} for a
closed total path \cite{Storey,Borde08}.  As a specific example, let
us consider the prediction from GR, which has been extensively treated
in \cite{Borde01}; here we only present a very basic analysis
sufficient for our purposes. The appropriate Lagrangian is given by
the proper time $\textrm{d}\tau=(-g_{\mu\nu}\textrm{d} x^\mu
\textrm{d} x^\nu/c^2)^{1/2}$, i.e. $L_\textrm{GR}=
-mc^2\textrm{d}\tau/\textrm{d}t$, and is derived to sufficient
accuracy using the Schwarzschild metric generated by the Earth,
\begin{equation}
  L_\textrm{GR}(z,\dot{z})=-mc^2+\frac{GMm}{r_\oplus}
  -mgz+\frac{1}{2}m\dot{z}^2
  \,, \label{LGR}
\end{equation}
where $r_\oplus$ is the Earth's radius, $g={GM}/{r_\oplus^2}$ is the
Newtonian gravitational acceleration, $G$ is Newton's gravitational
constant, $M$ is the mass of the Earth, $m$ the mass of the atom, $c$
the speed of light in vacuum, $z$ is defined by $r=r_\oplus+z$ with
$r$ the radial coordinate, and we neglect the post-Newtonian
corrections. For simplicity we restrict ourselves to only radial
motion, which is sufficient for the arguments in this paper.

The equations of motion are deduced from \eqref{LGR} using the
principle of least action or the Euler-Lagrange equations, and read
evidently $\ddot{z}=-g$. Then, when integrating \eqref{LGR} along the
resulting paths, and calculating the difference of action integrals 
$\Delta \varphi_S$ as defined by
the first term in \eqref{phase_interf}, one finds
\begin{equation}\label{DeltaphiS}
\Delta \varphi_S = 0\,.
\end{equation}

The key point about the result \eqref{DeltaphiS} is a consistent
calculation of the two paths in the atom interferometer and of the
phases along these paths, both derived from the same classical action,
using in a standard way the principle of least action. At the deepest
level, the principle of least action and its use in atom
interferometry comes from the Feynman path integral formulation of
quantum mechanics or equivalently the Schr\"odinger equation
\cite{Storey}. Thus only the second term $\Delta \varphi_\ell$ remains
in \eqref{phase_interf}. The final phase shift,
\begin{equation}\label{phase_total}
\Delta \varphi = \Delta \varphi_\ell = k \,g \,T^2\,,
\end{equation}
depends on the wavevector $k$ of the lasers, on the interrogation
time $T$ and on the local gravity $g$.
This shows that the atom interferometer is a gravimeter or
accelerometer. The phase shift \eqref{phase_total} arises entirely
from the interactions with the lasers and the fact that the atoms are
falling with respect to the laboratory in which the experiment is
performed. The atom's Compton frequency is irrelevant.

\subsection{Analysis in the modified Lagrangian
formalism}\label{modified}

This framework, which we call the ``modified Lagrangian formalism'',
is an adaptation for our purpose of a powerful formalism for analyzing
tests of the EEP and its various facets: the WEP, the local Lorentz
invariance (LLI) and the local position invariance (LPI)
\cite{Nordtvedt,Haugan,Will}. This formalism allows deviations from
general relativity and metric theories of gravity, with violations of
the UFF and UCR, and permits a coherent analysis of atom
interferometry experiments.

The formalism is defined by a single Lagrangian, that is however
different from the GR Lagrangian. For our purposes it is sufficient to
use a strongly simplified ``toy'' Lagrangian chosen as a particular
case within the ``energy conservation formalism'' of Nordtvedt and
Haugan \cite{Nordtvedt,Haugan} (see \cite{Will} for a review). To keep
in line with MPC we choose an expression similar to the Lagrangian of
GR given by \eqref{LGR}, namely
\begin{equation}\label{Lmodified}
L_\textrm{modified} = - m_0 \,c^2+\frac{GM m_0}{r_\oplus} -
\bigl(1+\beta^{(a)}_X\bigr) \,m_0 \,g\,z +\frac{1}{2}m_0\,\dot{z}^2\,,
\end{equation}
where $\beta^{(a)}_X$ denotes a dimensionless parameter characterizing
the violation of LPI, and depending on the particular type $X$ of 
mass-energy or interaction under consideration; e.g. $\beta^{(a)}_X$
would be different for the electromagnetic or the nuclear
interactions, with possible variations as a function of spin or the
other internal properties of the atom, here labelled by the
superscript $(a)$. Thus $\beta^{(a)}_X$ would depend not only on the
type of internal energy $X$ but also on the type of atom
$(a)$.

The Lagrangian \eqref{Lmodified} describes the Newtonian limit of a
large class of non-metric theories, in a way consistent with Schiff's
conjecture and fundamental principles of quantum mechanics. Most
alternative theories commonly considered belong to this class which
encompasses a large number of models and frameworks (see \cite{Will}
and references therein), like most non-metric theories
(e.g. the Belinfante-Swihart theory \cite{BelinfanteS}), some models
motivated by string theory \cite{DamourP}, some general parameterized
frameworks like the energy conservation formalism
\cite{Nordtvedt,Haugan}, the $\textrm{TH}\varepsilon\mu$ formalism
\cite{LightLee}, and the Lorentz violating standard model extension
(SME) \cite{Koste1,Koste2}.

Within this general formalism there is no fundamental distinction
between UFF and UCR tests, as violation of one implies violation of
the other, thus testing one implies testing the other. Depending on
the theory used, different experiments test different parameters or
parameter combinations at differing accuracies, so they may be
complementary or redundant depending on the context. By varying
\eqref{Lmodified} we obtain the equations of motion of the atom as
\begin{equation}\label{acc}
\ddot{z} = - \bigl(1+\beta^{(a)}_X\bigr)\,g\,,
\end{equation}
which shows that the trajectory of the atom is affected by the
violation of LPI and is not universal. In fact we see that
$\beta^{(a)}_X$ measures the non-universality of the ratio between the
atom's passive gravitational mass and inertial mass. Thus, in the
modified Lagrangian framework the violation of LPI implies a violation
of WEP and the UFF, and $\beta^{(a)}_X$ appears to be the
UFF-violating parameter. This is a classic example
\cite{Nordtvedt,Haugan} of the validity of Schiff's conjecture, namely
that it is impossible in any consistent theory of gravity to violate
LPI (or LLI) without also violating WEP.

The violation of LPI is best reflected in classical redshift
experiments with clocks which can be analysed using a cyclic gedanken
experiment based on energy conservation
\cite{Nordtvedt,Haugan,Will}. The result for the frequency shift in a
Pound-Rebka type experiment is
\begin{equation}\label{Zalpha}
Z = \bigl(1+\alpha^{(a)}_X\bigr)\,\frac{g \,\Delta z}{c^2}\,,
\end{equation}
where the redshift violating (or UCR violating) parameter
$\alpha^{(a)}_X$ is again non-universal. The important
point, proven in \cite{Nordtvedt,Haugan,Will}, is that
the UCR-violating parameter $\alpha^{(a)}_X$ is related in a
precise way to the UFF-violating parameter $\beta^{(a)}_X$, namely
\begin{equation}\label{alphabeta}
\beta^{(a)}_X = \alpha^{(a)}_X\,\frac{\overline{E}_X}{\overline{m} \,c^2}\,,
\end{equation}
where $\overline{E}_X$ is the internal energy responsible for the
violation of LPI, and $\overline{m}$ is the sum of the rest masses of
the particles constituting the atom.  Therefore we can compare the
different qualitative meaning of tests of UCR and UFF. For a given set
of UFF and UCR tests their relative merit is given by
\eqref{alphabeta} and is dependent on the model used, i.e. the type of
anomalous energy $\overline{E}_X$ and its dependence on the used
materials or atoms.

We now consider the application to atom interferometry. In the
experiment of \cite{Peters,Mueller} the ``atom-clock'' that
accumulates a phase is of identical composition to the falling object
(the same atom), hence one has to consistently use the same value of
$\beta^{(a)}_X$ when calculating the trajectories and the phase
difference using the Lagrangian \eqref{Lmodified}. It is then easy to
show that $\Delta\varphi_S=0$ with the above Lagrangian
\cite{Storey,Borde01,Wolf}. The vanishing of $\Delta\varphi_S$ in this
case is a general property of all quadratic Lagrangians and comes from
consistently using the same Lagrangian for the calculation of the
trajectories and the phase shift. It is related to the cancellation
between the kinetic term and the gravitational potential energy term
in the Lagrangian \eqref{Lmodified}.

Then the total phase shift of the atom interferometer is again given
by the light interactions only, which are obtained from the phases at
the interaction points evaluated using the trajectory given by
\eqref{Lmodified}. One then obtains
\begin{equation}\label{dphibetaX}
\Delta\varphi = \Delta\varphi_\ell =
\bigl(1+\beta^{(a)}_X\bigr)k\,g\,T^2\,.
\end{equation}
We first note that in this class of theories the Compton frequency
plays no role, as $\Delta\varphi_S = 0$. Second, we note that although
$\beta^{(a)}_X$ appears in the final phase shift, this is entirely
related to the light phase shift coming from the trajectory of the
atoms, and thus is a measurement of the effective free fall
acceleration $(1+\beta^{(a)}_X)g$ of the atoms, which is given by
\eqref{acc}. In \cite{Peters,Mueller} the resulting phase shift is
compared to $k\,\tilde{g}\,T^2$ where $\tilde{g}$ is the measured free
fall acceleration of a falling macroscopic corner cube,
i.e. $\tilde{g}=\bigl[1+\beta^{(\textrm{corner cube})}_X\bigr]g$, also
deduced from \eqref{acc}. In this class of theories the experiment is
thus a test of the UFF, as it measures the differential gravitational
acceleration of two test masses (C{\ae}sium atom and corner cube) of
different internal composition, with precision
\begin{equation}\label{testUFF}
\left|\beta^{(\textrm{Cs})}_X - \beta^{(\textrm{corner cube})}_X
\right|\lesssim 7 \times 10^{-9}\,.
\end{equation}
Note that this expression is
equivalent to the one obtained by MPC in their recent paper
\cite{Hohensee2} (p.~4), but different from the one obtained in
the same paper (p.~3) for UCR tests.\footnote{In \cite{Hohensee2}
atom interferometer tests set limits on the same parameter
combination as classical UFF tests, namely $\beta_1+\xi^{\rm
bind}_1-\beta_2-\xi^{\rm bind}_2$ (\cite{Hohensee2}, p.~4), where
the subscripts refer to test masses 1 and 2 (e.g. $1=$ Cs and
$2=$ falling corner cube in \cite{Peters} whilst $1=$ Ti and $2=$ Be
in \cite{Schlamminger}), and $\xi^{\rm bind}$ refers to the
nuclear binding energy of the test masses. On the other hand UCR
tests in \cite{Hohensee2} set limits on either $\beta_1-\xi^{\rm
trans}_2$ or $\xi^{\rm trans}_1-\xi^{\rm trans}_2$
(\cite{Hohensee2}, p.~3), where $\xi^{\rm trans}$ is related to
the atomic transition of the atom used in the clock and not to its
nuclear binding energy.}

\subsection{Analysis in the multiple Lagrangian
formalism}\label{multiple}

In the second framework, which we call the ``multiple Lagrangian
formalism'', the \textit{motion} of test particles (atoms or
macroscopic bodies) obeys the standard GR Lagrangian in a
gravitational field, whereas the \textit{phase} of the corresponding
matter waves obeys a \textit{different} Lagrangian.  The MPC analysis
in \cite{Mueller} belongs to this framework,\footnote{See ``Methods''
  in \cite{Mueller}, where they consider two scenarios. The first one
  clearly states that the trajectories are not modified whilst the
  atomic phases are. The second uses $g'$ for the trajectories and
  $g(1+\beta)$ for the phases, which again corresponds to two
  different Lagrangians.}  which raises extremely difficult problems.

The WEP is assumed to be valid for the motion of massive classical
particles which thus obeys the standard Lagrangian of general
relativity, $L_\textrm{particle} = L_\textrm{GR}$, i.e. to first order
\begin{equation}\label{Lpart}
L_\textrm{particle} = -m\,c^2+\frac{GMm}{r_\oplus}-m\,g\,z+\frac{1}{2}m\dot{z}^2\,.
\end{equation}
On the other hand the action integral to be used for the computation
of the phase shift of the quantum matter wave is assumed to be
calculated from the different Lagrangian
\begin{equation}\label{Lmueller}
L_\textrm{wave} =-m\,c^2+\frac{GMm}{r_\oplus}-(1+\beta)m\,g\,z
+\frac{1}{2}m \dot{z}^2\,,
\end{equation}
where the parameter $\beta$ that measures the deviation from GR
(i.e. $\beta=0$ in GR) enters as a correction in the atom's
gravitational potential energy. Integrating the Lagrangian \eqref{Lmueller} along the paths given by the Lagrangian \eqref{Lpart} shows that $\Delta\varphi_S = \beta \,k \,g \,T^2$, where $\beta$ takes the meaning of a redshift-violation
parameter. In particular we notice that $\beta$ could be ``universal'', in contrast with the
parameter $\beta^{(a)}_X$ in \eqref{Lmodified} which depends on the
type of atom $(a)$ and on some internal energy $X$ violating LPI. In
the multiple Lagrangian formalism, because WEP is valid we can always
test the value of $\beta$ by comparing $g$ as obtained from the free-fall of test
bodies, to the phase difference cumulated by matter waves (as proposed by MPC \cite{Mueller}).

The most important problem of this formalism is that it is
inconsistent to use a different Lagrangian (or metric) for the
calculation of the trajectories and for the phases of the atoms or
clocks. More precisely, it supposes that the fundamental Feynman path
integral formulation of quantum mechanics, which is at the basis of
the derivation of the phase shift in an atom interferometer
\cite{Storey}, has to be altered in the presence of a gravitational
field or could be wrong. Physically it amounts to making the
distinction between the atoms when calculating their trajectories and
the same atoms when calculating their phases, which is inconsistent as
it is the same fundamental matter field in both cases. Even more, the
basis for the derivation of the atom interferometry phase shifts is
unjustified because the Feynman formalism is violated. To remain
coherent some alternative formalism for the atomic phase shift
calculations (presumably modifying quantum mechanics) should be
developed and used. More generally the multiple Lagrangian formalism
supposes that the duality between particles and waves in quantum
mechanics gets somehow violated in a gravitational field.

The above violation of the ``particle-wave duality'', implied by the
multiple Lagrangian formalism, is very different from a violation of
the equivalence principle in the ordinary sense. In this formalism a
single physical object, the atom, is assumed to be described by two
different Lagrangians, $L_\text{wave}$ applying to its phase shift,
and $L_\text{particle}$ applying to its trajectory. By contrast, in
tests of the equivalence principle, one looks for the modification of
the free fall trajectories or clock rates as a function of composition
or clock type. Thus, different bodies or clocks are described by
different Lagrangians, but of course for any single type of body we
always have $L_\text{wave}=L_\text{particle}$. The pertinent test-bed
for equivalence principle violations is the modified Lagrangian
formalism reviewed in Section \ref{modified}, which does not imply any
particle-wave duality violation.

The second problem of theories in the multiple Lagrangian formalism
and of the interpretation of MPC is the violation of Schiff's
conjecture \cite{Schiff}. Indeed, we have assumed in this Section that
the LPI aspect of the equivalence principle is violated but that for
instance the WEP aspect remains satisfied. In a complete and
self-consistent theory of gravitation one expects that the three
aspects of the Einstein equivalence principle (WEP, LLI and LPI) are
sufficiently entangled together by the mathematical formalism of the
theory that it is impossible to violate one without violating all of
them. The Schiff conjecture has been proved using general arguments
based upon the assumption of energy conservation
\cite{Nordtvedt,Haugan}; it is satisfied, for instance, in the
$\textrm{TH}\varepsilon\mu$ formalism \cite{LightLee}.  A contrario,
we expect that violating the conjecture leads to some breakdown of
energy conservation. In the present case this translates into
postulating two Lagrangians \eqref{Lpart} and \eqref{Lmueller} for the
same physical object. Explicit theories that are logically and
mathematically consistent but still violate Schiff's conjecture are
very uncommon (see \cite{Will} and references therein).

We conclude that the general statement of MPC according to which the
atom interferometer measures or tests the gravitational redshift at
the Compton frequency is incorrect. Instead, the interpretation of the
experiment needs to be considered in the light of alternative theories
or frameworks. Although one can consider particular alternative
frameworks in which the statement of MPC could make sense, such
frameworks raise unacceptable conceptual problems which are not at the
moment treated in a satisfactory manner. In particular, they break the
fundamental principles of quantum mechanics which are used for
calculating matter wave phases. In most common and plausible
theoretical frameworks the atom interferometry experiment tests the
universality of free fall with the Compton frequency being irrelevant.

\end{document}